\documentclass[preprint]{iucr}              

\usepackage[bottom]{footmisc}
\usepackage{amssymb} 
\usepackage{url} 
\usepackage{graphicx}
\usepackage{grffile}
\usepackage{array}
\usepackage{rotating} 
\usepackage{varwidth} 
\usepackage{nameref}
\usepackage{ulem} 

\usepackage[usenames, dvipsnames]{xcolor}
\usepackage{cancel}
\usepackage{amsmath}

\newcommand{\Eq}[1]{Eq.~(\ref{#1})}

\let\foo\caption

\ifPDF
  \PassOptionsToPackage{pdftex,bookmarksopen,bookmarksnumbered}{hyperref}
  \RequirePackage{hyperref}
\fi
\let\caption\foo

\providecommand{\href}[2]{#2} 

\ifPDF
  \RequirePackage{hyperref}
  \PassOptionsToPackage{pdftex,bookmarksopen,bookmarksnumbered}{hyperref}
\fi
\let\caption\foo

\paperprodcode{a000000}      
\paperref{xx9999}            
\papertype{IU}               
\paperlang{english}          

\journalyr{2003}
\journalreceived{\relax}
\journalaccepted{\relax}
\journalonline{\relax}

\begin{document}                  

\title{Fast Eikonal Phase Retrieval for High-Throughput Beamlines}

\author[a]{Alessandro}{Mirone}{ mirone@esrf.fr}
\author[a,b]{Theresa}{Urban}
\author[a,b]{Joseph}{Brunet}
\author[b]{Claire L.}{Walsh}
\author[b]{Peter D.}{Lee}
\author[a]{Paul}{Tafforeau}

\aff[a]{European Synchrotron Radiation Facility, 71, Avenue des Martyrs, Grenoble F-38000. \country{France}}
\aff[b]{Department of Mechanical Engineering, University College London, London, UK.}

\shortauthor{Mirone et al.}

\begin{synopsis}
Fast eikonal phase retrieval (EPR) with complementary local (sub-pixel) and non-local (multi-pixel) solvers strongly suppresses nonlinear streak artefacts in long-distance PPC-$\mu$CT at high computational efficiency.
\end{synopsis}

\begin{abstract}
  We introduce a fast Eikonal Phase Retrieval (EPR) formulation that accelerates eikonal phase retrieval by more than two orders of magnitude while retaining controlled accuracy. The method is derived from a second-order asymptotic expansion in the propagation distance $L$ and complemented by the leading Wentzel--Kramers--Brillouin (WKB)
  wave-optics correction, yielding an efficient iterative correction scheme preconditioned by FFT-diagonal, energy-dependent inverse operators (Paganin-type filters).
To ensure robustness across practical experimental regimes, we combine two complementary solvers:
(i) a local $O(L^2)$ closure that is accurate when eikonal shifts remain sub-pixel, and
(ii) a non-local formulation for multi-pixel shifts, in which intensity is propagated through an explicit eikonal ray mapping using a mass-conserving bilinear redisribution on the detector grid, and detector residuals are transferred back to the object grid by the corresponding adjoint (transpose), implemented as bilinear interpolation, before applying an approximate FFT-diagonal preconditioner to accelerate convergence. The same framework supports polychromatic data through a compact spectral discretisation, allowing energy-dependent transport and inversion while keeping the iteration GPU/FFT efficient. Overall, this unified approach enables accurate and computationally efficient phase retrieval across propagation conditions relevant to high-throughput PPC-$\mu$CT experiments.
\end{abstract}



\section{Introduction}

Propagation-based phase-contrast micro-tomography (PPC-$\mu$CT) is one of the most widely used imaging modalities at synchrotron facilities, owing to its experimental simplicity, high photon efficiency and sensitivity across a broad range of applications. In its simplest implementation, provided that sufficient transverse coherence is available, PPC-$\mu$CT requires only moving the detector downstream with respect to a conventional absorption radiography set-up, which makes it particularly well suited to high-throughput imaging workflows.

The advent of fourth-generation synchrotron sources has increased beam brilliance while reducing the effective source size. As a consequence, longer sample-to-detector propagation distances can be exploited without incurring significant geometrical blur from the finite source size (circle of confusion). This enables operation in a propagation-enhanced near-field regime, where phase-contrast sensitivity is strongly increased and weakly refracting features can become visible above statistical noise. This regime is now routinely exploited on high-energy, high-throughput beamlines such as BM18 at the ESRF, notably in the context of hierarchical phase-contrast tomography (HiP-CT)~\cite{Walsh2021}.

However, the gain in sensitivity comes with important challenges. In realistic experiments on large and heterogeneous specimens, weakly refracting structures coexist with strongly absorbing regions and sharp interfaces, such as bone--air or bone--fluid boundaries. At increased propagation distances, the intensity modulations generated by steep phase gradients at such interfaces are strongly amplified. Under these conditions, the linearised assumptions underlying standard single-distance phase retrieval~\cite{pag2002} break down, leading to characteristic nonlinear artefacts~\cite{Mohan2020,Mirone2025JSR,Urban2025HiPCT}, including long-range streaks that severely degrade image quality and compromise downstream analysis.

A Fresnel wave-optics formulation~\cite{Mohan2020} can, in principle, model near-field propagation even under strong refraction, but it requires an explicit pixel-wise representation of the complex wavefield. This introduces strict sampling constraints: when phase variations become too rapid to be sampled at the detector pitch, the discretised wavefield is no longer adequately band-limited and numerical Fresnel propagation becomes contaminated by aliasing unless very large oversampling factors are used. In propagation-enhanced PPC-$\mu$CT on heterogeneous specimens, sharp interfaces naturally produce such rapid phase variations, making brute-force Fresnel propagation quickly impractical in terms of memory footprint and computational cost.

By contrast, Paganin-type single-distance transport-of-intensity approaches~\cite{pag2002} do not require an explicit pixel-wise phase map. Propagation is modelled through intensity transport under a linearised phase-gradient assumption, and the computation remains numerically stable even in situations where a fully sampled wavefield representation would be impractical. Their limitation is therefore primarily physical rather than numerical: once phase gradients and intensity dynamics become sufficiently strong, the linearisation fails and nonlinear streak artefacts appear.

In our previous work we introduced the Eikonal Phase Retrieval (EPR) algorithm, an iterative phase-retrieval framework based on the eikonal approximation~\cite{Mirone2025JSR}. EPR was designed to go beyond the linearised transport-of-intensity description and to remain valid in the presence of strong phase gradients and large intensity dynamics, where direct Fresnel propagation becomes computationally impractical. Extensive experimental validation on challenging datasets, including strongly absorbing biological samples acquired on BM18, demonstrated that EPR suppresses or strongly reduces linearisation artefacts and recovers fine structural details that are otherwise lost with standard linear approaches~\cite{Mohan2020,Mirone2025JSR}. In this sense, EPR established that a more faithful physical model of near-field propagation is essential in the propagation-enhanced phase-contrast regime to reach optimal sensitivity.

Despite these advantages, the first implementation of EPR suffered from a major practical limitation: computational cost. For large HiP-CT datasets, the full EPR workflow could require up to several tens of hours on multi-GPU systems, whereas the experimental throughput of the beamline is typically of the order of one sample every few hours. This mismatch between reconstruction time and acquisition rate has so far hindered routine use.

Motivated by this limitation, several strategies were explored to reduce the computational burden. In particular, preliminary work investigated small convolutional neural networks to identify or correct the regions of the projections most responsible for nonlinear artefacts. Remarkably, these networks relied exclusively on local quantities, such as intensity values in a neighbourhood of each pixel and their planar derivatives, yet achieved promising results with relatively fast training~\cite{Admans2024}. This observation suggests that, over a substantial fraction of the experimental regime of interest, the underlying physics is only weakly nonlinear and dominated by correction terms that admit a local differential representation.

This insight directly motivates the present work. We introduce a fast EPR formulation based on a reduced semi-analytic near-field model and an efficient FFT-preconditioned iterative inversion. In particular, we (i) retain the complete $O(L^2)$ content of a WKB (saddle-point) expansion of the Fresnel propagator, (ii) complement the local second-order closure with a non-local mapping-based solver robust to multi-pixel shifts, (iii) support polychromatic data through a compact spectral discretisation, and (iv) achieve runtimes compatible with high-throughput reconstruction workflows while strongly suppressing nonlinear streak artefacts.

Section~\ref{sec:principle} summarises the principle and algorithm at a level sufficient to interpret the Results. The detailed formalism is provided in the Methods, while full derivations and additional implementation details are placed in the Supporting Information.

\section{Principle and algorithm overview}
\label{sec:principle}

This section summarises the forward model and inversion scheme used throughout the paper, introducing the notation needed to interpret the Results. Full derivations and additional implementation details are provided in the Methods and Supporting Information.

\paragraph{Forward model and WKB terminology.}
We describe near-field propagation using a Wentzel--Kramers--Brillouin (WKB)~\cite{Wentzel1926WKB,BenderOrszag1999}
expansion of the Fresnel propagator.
In this hierarchy, \emph{WKB0} corresponds to the leading eikonal (geometrical-optics) transport of intensity along rays,
while \emph{WKB1} represents the first wave-optics correction (diffraction-pressure / quantum-potential term).
Accordingly, the detector intensity can be expressed as a linearised transport term plus higher-order corrections in the
propagation distance $L$.

While higher-order terms can in principle be derived, their physical interpretation becomes progressively less transparent
as one moves away from the near-field, single-valued-ray regime.
Beyond the leading eikonal transport (WKB0), higher-order contributions increasingly encode wavefront physics that cannot
be fully captured by a local, point-to-point WKB ray mapping.
For this reason, we retain WKB0 and include only the leading WKB1 correction at $O(L^2)$ as a compact estimate of
wave-optics effects within this near-field WKB framework.

\paragraph{Linear term and relation to Paganin.}
The $O(L)$ term is the standard linearised transport-of-intensity description and, under the homogeneous-object assumption, its FFT-diagonal inverse reduces to the well-known single-distance Paganin filter~\cite{pag2002}. In the iterative algorithm, this inverse is used as a \emph{preconditioner}, meaning that we apply it to the back-transported residual as an approximate inverse to accelerate convergence. We denote by $\mathcal{P}_{1,s}^{-1}$ the FFT-diagonal approximate inverse of the linearised ($O(L)$) per-line operator (Paganin-type filter), and use it as a preconditioner.

\paragraph{Eikonal transverse shift and local versus non-local regime.}
Refraction induces a transverse ray displacement
$\boldsymbol{\Delta}(\boldsymbol r) = \frac{L}{k}\nabla_{\!\perp}\phi(\boldsymbol r)$,
where $\phi(\boldsymbol r)$ is the object-plane phase at $z=0$ and $k=2\pi/\lambda$.
We characterise the regime by the shift-to-pixel ratio
$\eta(\boldsymbol r)=\|\boldsymbol{\Delta}(\boldsymbol r)\|/p$,
where $p$ is the detector pixel size (or the internal step on an oversampled grid).
When $\eta < 1$ over most of the field of view, the propagated intensity can be expanded locally in powers of $L$, leading to an efficient $O(L^2)$ differential forward model.
When $\eta > 1$, the forward physics becomes effectively non-local: a detector pixel receives its dominant contribution from object-plane locations displaced by one or more pixels. In that regime, we retain the explicit WKB0 ray mapping and evaluate the forward operator numerically as a conservative bilinear redistribution on the detector grid. The detector residual is then transferred back to the object grid with the corresponding adjoint (transpose) operator, implemented as bilinear interpolation at the mapped coordinates, after which the preconditioner  $\mathcal{P}_{1,s}^{-1}$  is applied to accelerate convergence.

\paragraph{Polychromatic data.}
Polychromatic measurements are modelled as an incoherent sum of $N_s$ spectral components. All per-line quantities (phase/absorption parameters and shifts) are treated energy-dependently while keeping the iteration FFT efficient. If the effective spectrum is spatially uniform, the model reduces to a single-spectrum description; if needed, it can also accommodate slow spatial variations of the effective spectrum across the detector (see Methods and Supporting Information).

\section{Results}

\subsection{Sheep-head HiP-CT dataset and region of interest}

Here we apply the local and non-local methods to the sheep-head HiP-CT dataset~\cite{Urban2025HiPCT}, previously used to introduce and validate the original Eikonal Phase Retrieval (EPR) method. The experiment follows the HiP-CT protocol, where the specimen is mounted in ethanol and acquired at long propagation distance to maximise propagation-based contrast on weakly absorbing soft tissues. This dataset is particularly challenging because the slice contains, simultaneously, (i) low-contrast anatomical structures (soft tissues) and (ii) very strong phase gradients and absorption variations associated with bones and bone interfaces, which generate pronounced propagation-induced nonlinear effects and long-range artefacts.

\begin{figure}
\centering
\includegraphics[width=\linewidth]{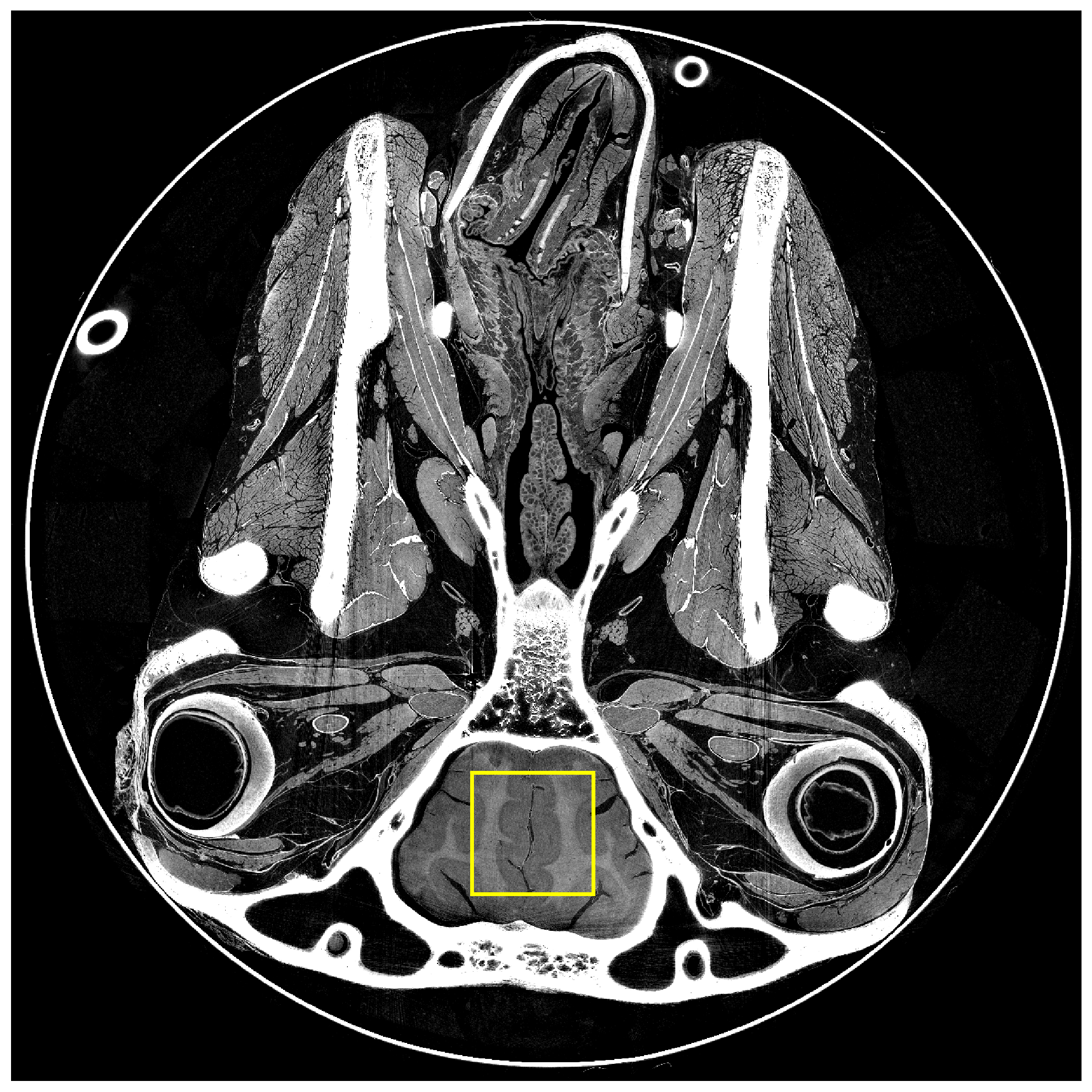}
\caption{Full reconstructed slice of the sheep-head HiP-CT dataset (same specimen and acquisition context as in the original EPR paper). The yellow rectangle marks the region of interest (ROI) used for the detailed comparisons in Figs.~\ref{figgridone}--\ref{figgridtwo}. Among the multiple zoomed regions discussed in the original EPR work, the ROI chosen here corresponds to the brain region (Inset \emph{A} in Fig.~3 of the original EPR paper), where soft-tissue contrast coexists with strong nearby gradients induced by adjacent bone structures.}
\label{figoverview}
\end{figure}

The ROI depicted in Fig.~\ref{figoverview} was selected to emphasise the regime most relevant to this work: the simultaneous presence of weak features and strong gradients, for which a purely linearised, single-distance forward model is known to produce characteristic artefacts. In particular, this brain-region ROI is representative because it contains fine anatomical texture in soft tissues, while remaining influenced by high-gradient structures (bones and interfaces) that drive the strongest propagation nonlinearities.

\subsection{Effect of the second-order ($L^{2}$) forward model and of polychromatic modelling}

\begin{figure}
\centering
\includegraphics[width=\linewidth]{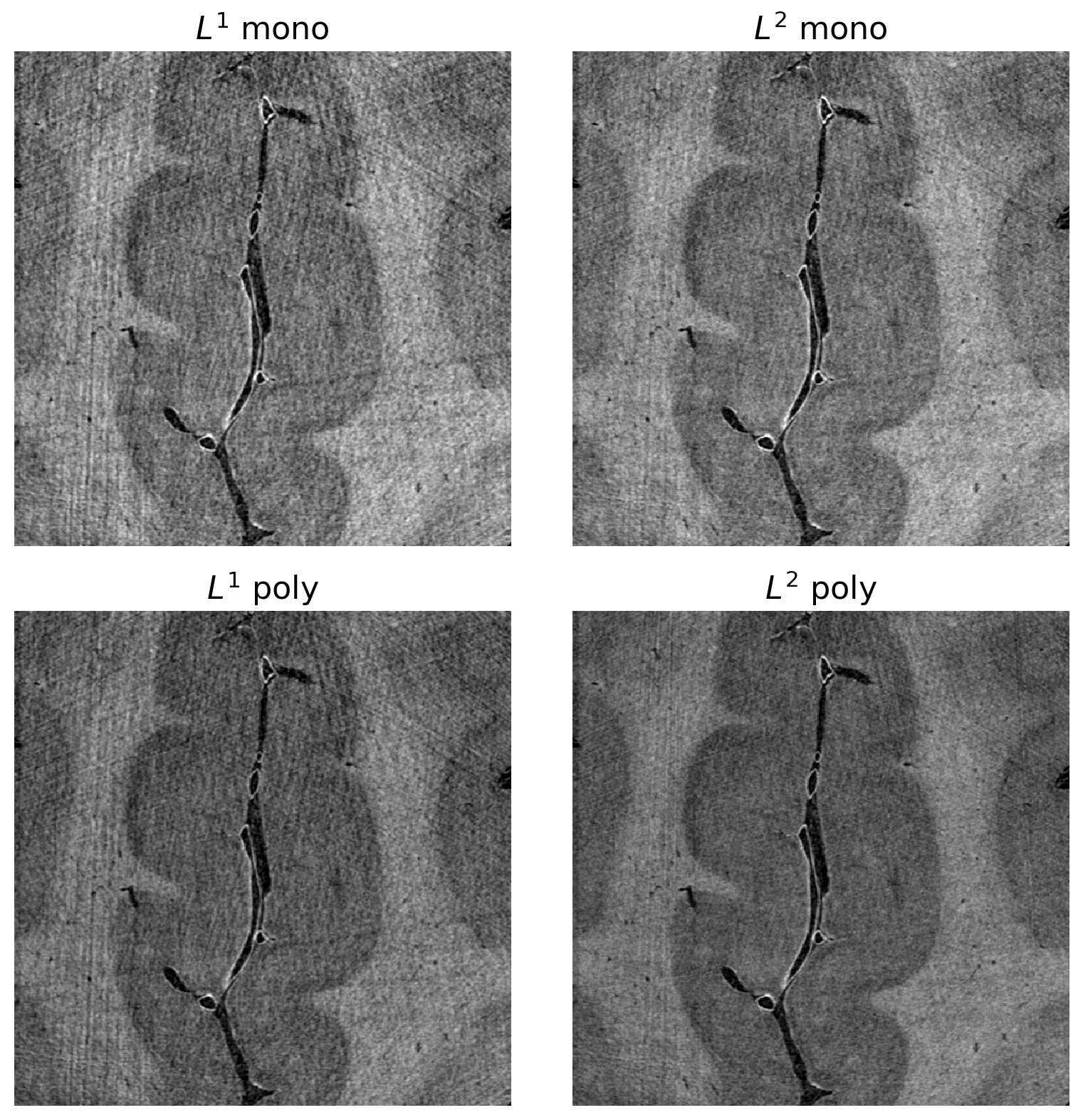}
\caption{ROI comparison (brain region) for the four forward models.
Top row: monochromatic models; bottom row: polychromatic models.
Left column: first-order model $L^{1}$; right column: second-order model $L^{2}$.
(Here $L^{1}$ mono is equivalent to the Paganin single-distance forward model in the homogeneous-object setting~\cite{pag2002}.)}
\label{figgridone}
\end{figure}

We compare four reconstructions of the same ROI, arranged in the $2\times 2$ grid of Fig.~\ref{figgridone}. The top row reports the monochromatic approximation (``mono''), while the bottom row reports the polychromatic model (``poly''). The left column corresponds to the first-order propagation model $L^{1}$, i.e.\ the forward (eikonal/WKB0 transport) truncated at $O(L)$, whereas the right column corresponds to the second-order model $L^{2}$, i.e.\ the forward truncated at $O(L^2)$ and including the additional diffraction-pressure (WKB1) contribution that enters the intensity at second order. (In the homogeneous-object setting, $L^{1}$ mono is equivalent to the standard Paganin phase-retrieval single-distance model~\cite{pag2002}.)

A first observation from Fig.~\ref{figgridone} is that the polychromatic modelling is consistently beneficial: the $L^{1}$ poly reconstruction is slightly improved with respect to $L^{1}$ mono, and the same trend is observed between $L^{2}$ poly and $L^{2}$ mono. This is expected, as the effective spectrum alters the mapping between absorption and phase shift and therefore the quantitative balance between attenuation and refraction in the forward model.

The dominant improvement, however, is obtained by moving from $L^{1}$ to $L^{2}$. Including the full second-order content (quadratic eikonal WKB0 transport plus the WKB1 diffraction-pressure correction) reduces the residual nonlinear propagation artefacts that persist under the first-order truncation, and provides a visibly cleaner rendering of soft-tissue contrast in the presence of nearby strong gradients. In practice, the $L^{2}$ models better suppress the characteristic propagation-induced distortions that remain when only the first-order transport term is retained.

\begin{figure}
\centering
\includegraphics[width=\linewidth]{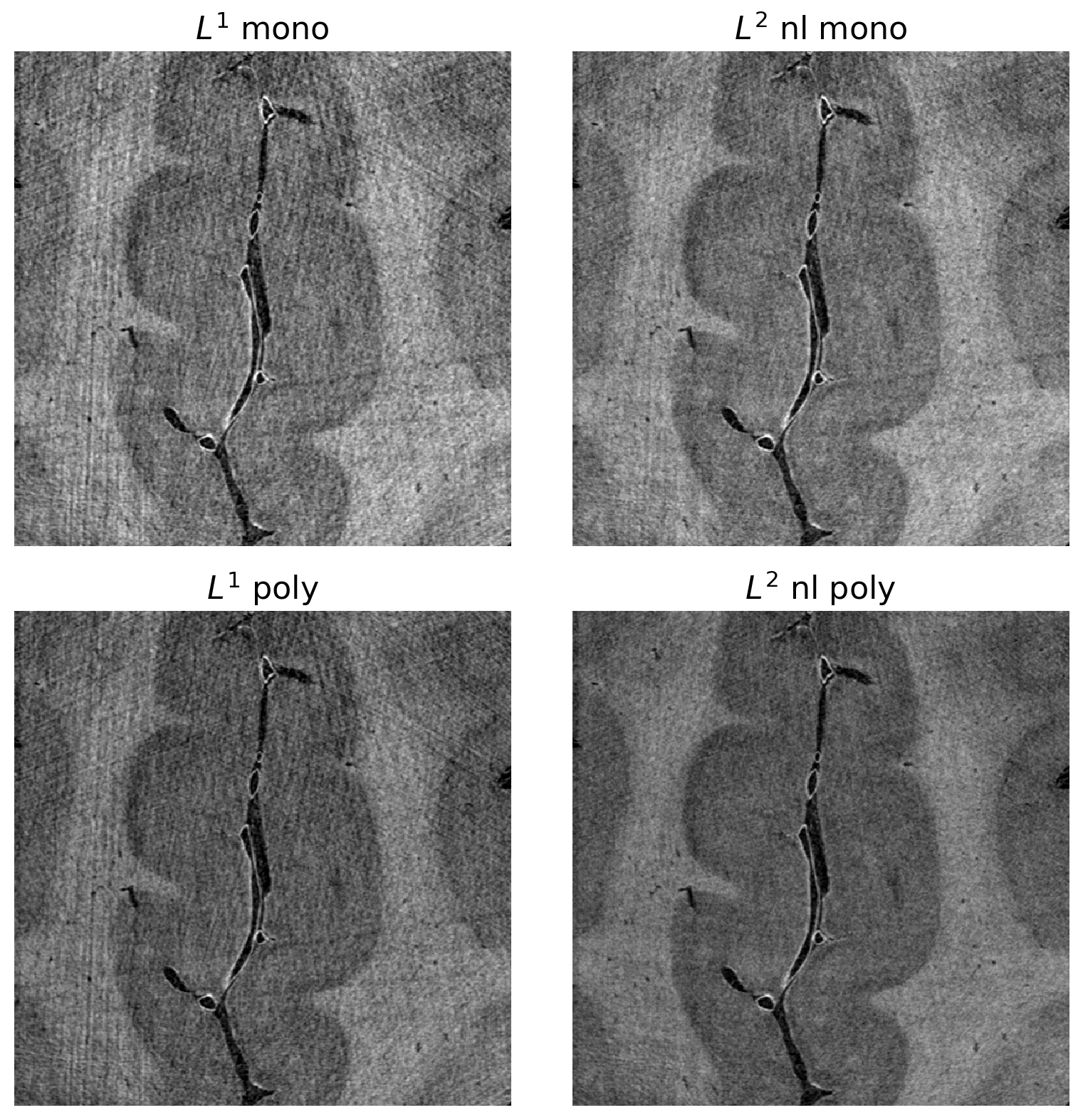}
\caption{Non-local: ROI comparison (brain region) for the four forward models.
Top row: monochromatic models; bottom row: polychromatic models.
Left column: first-order model $L^{1}$; right column: second-order non-local model $L^{2}$.
(Here $L^{1}$ mono is equivalent to the Paganin single-distance forward model in the homogeneous-object setting~\cite{pag2002}.)}
\label{figgridonenl}
\end{figure}

In Fig.~\ref{figgridonenl} the same comparison is repeated with the non-local solver, which remains stable and accurate when the estimated shift distribution extends beyond one pixel.

\begin{figure}
\centering
\includegraphics[width=\linewidth]{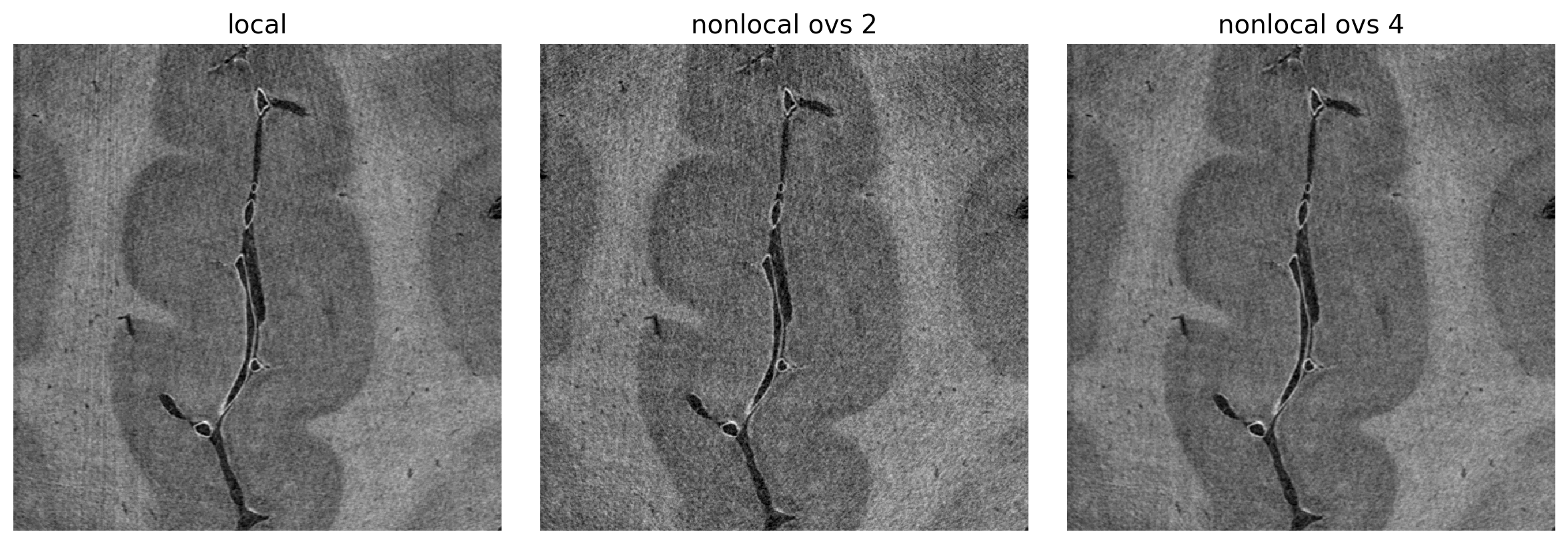}
\caption{Local vs non-local phase retrieval on the same ROI.
Left: local ($L^{2}$) solver (oversampling factor $2$).
Centre: non-local solver with oversampling factor $2$, showing residual high-frequency noise.
Right: non-local solver with oversampling factor $4$, where this noise is strongly reduced.}
\label{complocnonloc}
\end{figure}

All reconstructions reported in this section were computed on oversampled grids. Unless stated otherwise, the local solver uses an oversampling factor of $2$ (i.e.\ a pixel size reduced by a factor $2$), while the non-local solver uses an oversampling factor of $4$. Oversampling improves numerical robustness in two distinct ways. First, it reduces discretisation error in the transverse derivatives (gradients, Laplacians and divergences) used by the local $L^{2}$ forward model. 
Second, in the non-local solver, it reduces quantisation effects in the discrete ray-map transport implemented by bilinear redistribution on the detector grid and the corresponding discrete adjoint (transpose) interpolation back to the object grid, which can otherwise manifest as residual high-frequency texture when the displacement field spans a significant fraction of a pixel.
This effect is illustrated in Fig.~\ref{complocnonloc}, which compares the local solver (left) with the non-local solver at oversampling factors $2$ (centre) and $4$ (right). The residual high-frequency pattern visible in the non-local reconstruction at oversampling $2$ is strongly reduced when increasing the oversampling to $4$.

\subsection{Fast convergence of the $L^{2}$ polychromatic inversion}

\begin{figure}
\centering
\includegraphics[width=\linewidth]{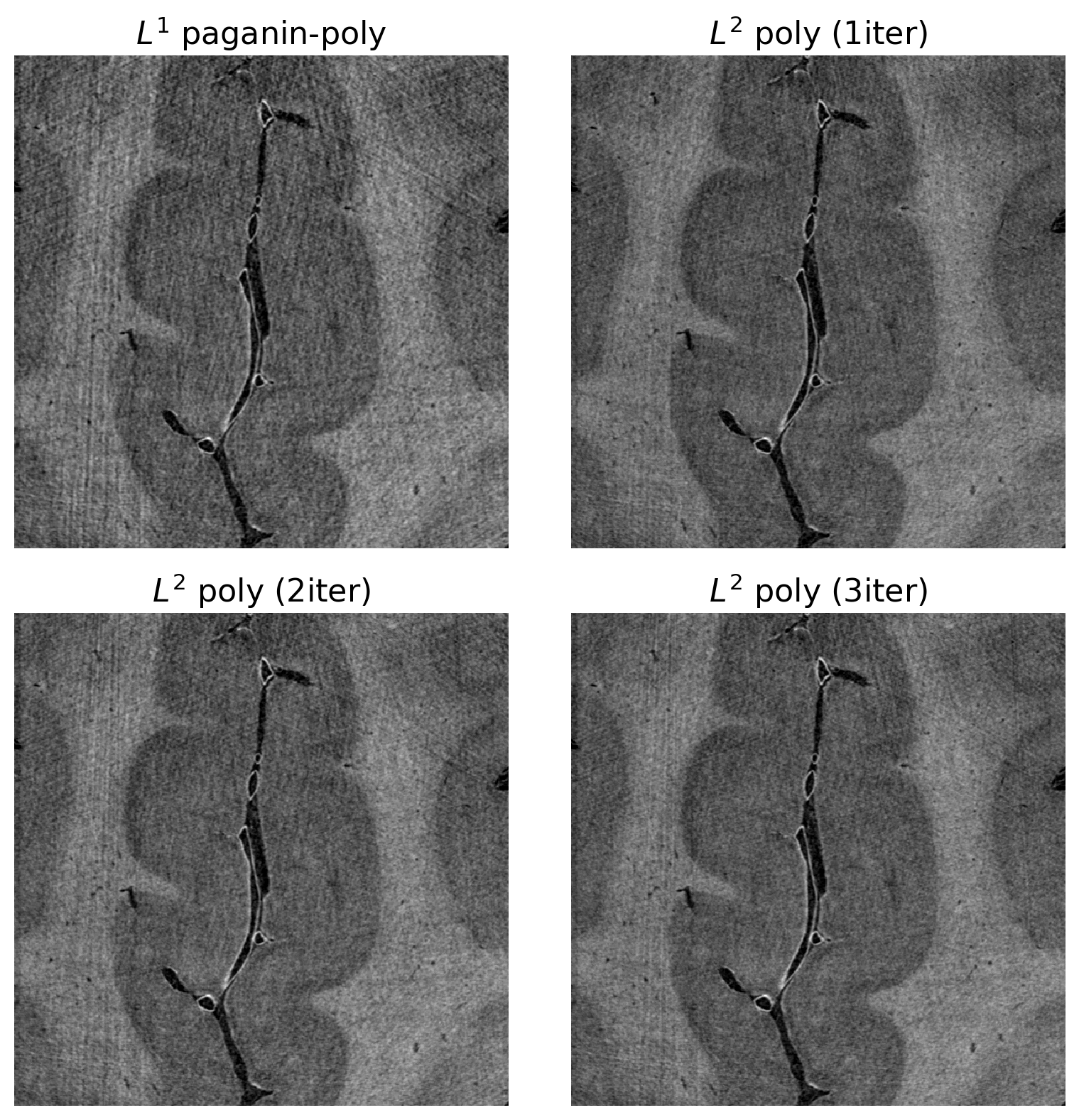}
\caption{Convergence in the polychromatic case.
A large improvement is already achieved when switching from $L^{1}$ poly to $L^{2}$ poly with a single iteration. Subsequent iterations produce only minor (often barely visible) refinements, indicating rapid convergence of the $L^{2}$ polychromatic inversion in this ROI.}
\label{figgridtwo}
\end{figure}

Fig.~\ref{figgridtwo} focuses on convergence in the polychromatic setting. For reference, the top-left panel reports the first-order solution ($L^{1}$ poly). The remaining panels show the second-order reconstruction ($L^{2}$ poly) after one, two, and three iterations. This makes it possible to assess both the improvement obtained when enabling the full $O(L^{2})$ forward physics and the marginal effect of additional iterations once the $L^{2}$ model is used.

The key result in Fig.~\ref{figgridtwo} is that the benefit of the second-order model appears immediately: going from $L^{1}$ poly to $L^{2}$ poly (1 iteration) already yields most of the visible improvement. Increasing the iteration count beyond the first $L^{2}$ update leads to changes that are comparatively subtle, suggesting that the algorithm converges very rapidly once the correct second-order physics is enabled in the forward model.

\subsection{Effect of the diffraction-pressure term}
Across the propagation distances and sample regimes explored in this work, the diffraction-pressure (WKB1) contribution is consistently negligible compared with the dominant WKB0 (eikonal transport) terms. We nevertheless include it in the forward model as the leading wave-optics correction in the same second-order expansion framework, and because it may become relevant in regimes closer to the boundary of validity of near-field approximations (e.g.\ at lower energy and/or higher spatial resolution), where diffraction effects can no longer be safely ignored.

\subsection{Robustness stress test beyond the sub-pixel regime: bamboo sample}
\label{sec:bamboo}

\begin{figure}
\centering
\includegraphics[width=\linewidth]{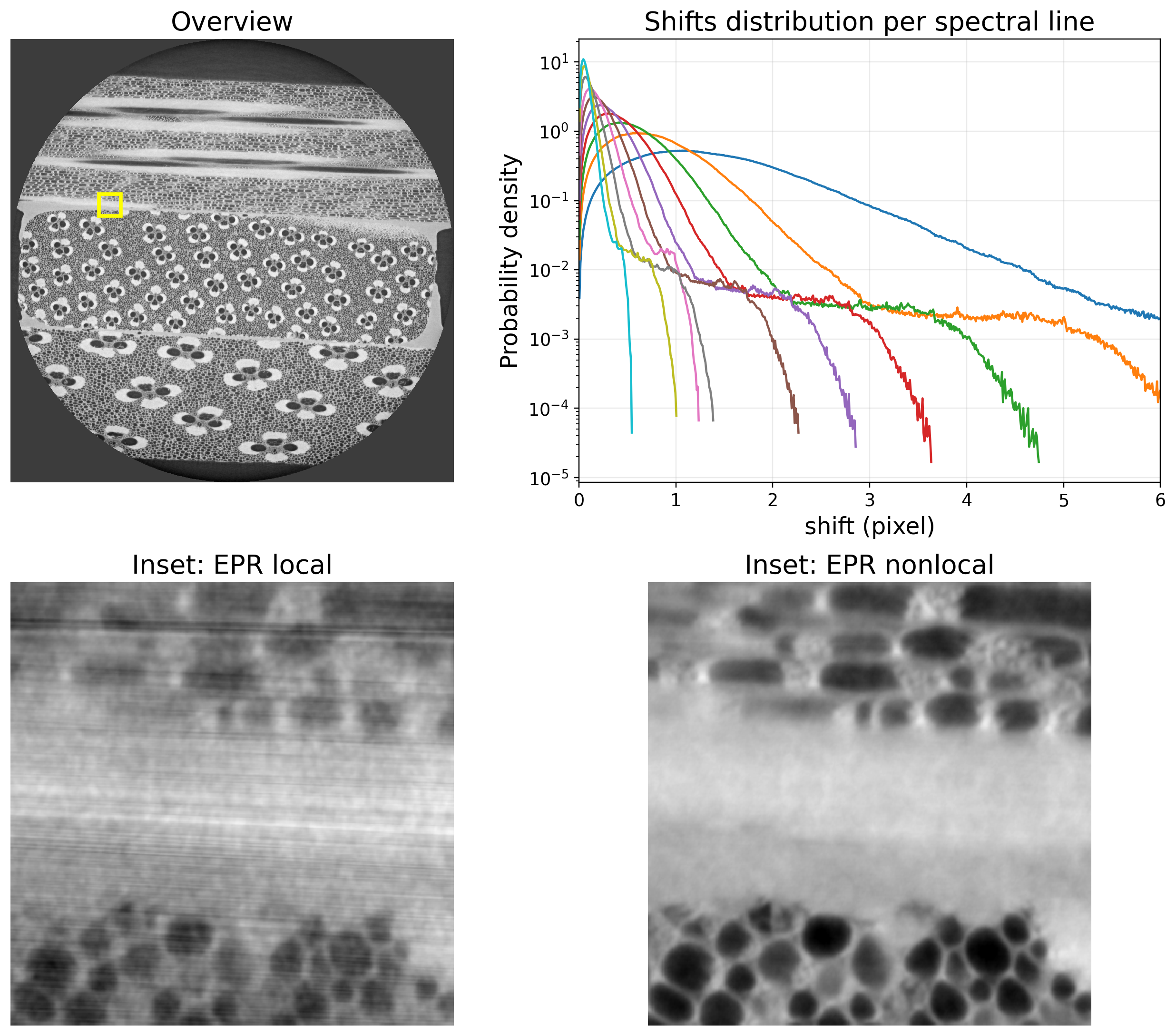}
\caption{
Zoom on a bamboo region acquired at $1.12~\mu\mathrm{m}$ voxel size, $71~\mathrm{keV}$ average beam energy, and $250~\mathrm{mm}$ propagation distance.
Top left: overview. Top right: distribution of estimated WKB0 transverse shifts (in detector-pixel units) for each sampling energy (spectral line) on a representative radiograph.
Bottom left: local solver after two iterations (diverging).
Bottom right: non-local solver with $10$ spectral lines, $5$ iterations.}
\label{bamboo}
\end{figure}

Figure~\ref{bamboo} reports a challenging bamboo sample acquired at $1.12~\mu\mathrm{m}$ voxel size, with an average beam energy of $71~\mathrm{keV}$ and a propagation distance of $250~\mathrm{mm}$. This long propagation distance was intentionally chosen, given the pixel size, as a stress test for the sub-pixel assumptions underlying the local solver.

The top-right panel of Fig.~\ref{bamboo} shows the distribution of estimated WKB0 transverse shifts (in detector-pixel units) for each sampling energy (spectral line) on a representative radiograph. A significant fraction of the field of view exhibits shifts larger than one pixel, placing this dataset outside the regime where a purely local $O(L^2)$ closure is expected to remain accurate and stable.
In this regime, the local solver fails (it degrades already at the first iteration and subsequently diverges), whereas the non-local solver remains stable. The bottom row of Fig.~\ref{bamboo} compares the local EPR result after two iterations (left) with the non-local result (right), obtained using $10$ spectral lines, $5$ iterations, and a relaxation factor of $0.4$. The divergence appears first where the strongest gradients are, along  the long longitudinal features.
The purpose of this bamboo dataset is to validate robustness when refraction-induced shifts exceed the detector pixel size. In this strong-shift regime, the mapping-based non-local solver remains stable, while the local $O(L^2)$ closure breaks down. In this particular stress-test we are already beyond the near-field limit where a single-valued eikonal, point-to-point mapping is justified. Extending the model beyond the eikonal setting is outside the scope of the present work and will be addressed in future developments.

\subsection{Computational performance}

We report indicative runtimes on a representative dataset comprising $8000$ radiographs of size $256\times 3104$ pixels (about $6.4\times 10^{9}$ pixels in total). All timings include radiograph reading (I/O) and the phase-retrieval computation, but exclude tomographic backprojection.

The benchmarks were obtained on a shared-memory server providing $96$ CPU cores (AMD EPYC 75F3 class) and one NVIDIA A40 GPU. The examples shown in this section can be reproduced by installing and running the \texttt{NightRail} workflow~\cite{Mirone2025NightRailBM18}, as detailed in the Supporting Information.

\begin{table}
\centering
\caption{Indicative runtimes for a dataset of $8000$ radiographs of size $256\times3104$ pixels.
All timings include radiograph reading (I/O) and phase retrieval, but exclude tomographic backprojection.
CPU timings were obtained on a $96$-core server (AMD EPYC 75F3 class), and GPU timings on one NVIDIA A40.}
\label{tab:runtimes}
\begin{tabular}{lccc}
\hline
Configuration & $N_s$ & Iter. & Time \\
\hline
\hline
$L^{1}$ mono (Paganin) on CPU & 1 & 1 & $45~\mathrm{s}$ \\

Local $L^{2}$ mono on GPU & 1 & 1 & $1.1~\mathrm{min}$ \\
Local $L^{2}$ mono on GPU & 1 & 2 & $1.23~\mathrm{min}$ \\
Local $L^{2}$ mono on CPU & 1 & 1 & $1.9~\mathrm{min}$ \\
Local $L^{2}$ mono on CPU & 1 & 2 & $2.7~\mathrm{min}$ \\

Local $L^{2}$ poly on GPU & 5 & 1 & $1.47~\mathrm{min}$ \\
Local $L^{2}$ poly on GPU & 5 & 2 & $1.9~\mathrm{min}$ \\
Local $L^{2}$ poly on CPU & 5 & 1 & $6.7~\mathrm{min}$ \\
Local $L^{2}$ poly on CPU & 5 & 2 & $11.3~\mathrm{min}$ \\

Non-local poly on GPU (oversampling $2$) & 5 & 1 & $2~\mathrm{min}$ \\
Non-local poly on GPU (oversampling $4$) & 5 & 1        & $8~\mathrm{min}$ \\
Original EPR implementation              & 5 & many     &  $870~\mathrm{min} $  \\
\hline
\end{tabular}
\end{table}

\paragraph{Baseline ($L^{1}$ mono, Paganin).}
The first-order monochromatic model ($L^{1}$ mono), equivalent to the standard Paganin single-distance retrieval under the homogeneous-object assumption, requires approximately $45~\mathrm{s}$ on the $96$-core CPU node.

\paragraph{Second order ($L^{2}$ mono).}
With the second-order model ($L^{2}$ mono), the runtime increases due to the additional planar-derivative evaluations required by the $O(L^{2})$ terms. For one iteration, the runtime is approximately $1.9~\mathrm{min}$ on CPUs and $1.1~\mathrm{min}$ on one GPU; for two iterations it is about $2.7~\mathrm{min}$ (CPU) and $1.23~\mathrm{min}$ (GPU).

\paragraph{Second order ($L^{2}$ poly).}
Using a five-line spectral discretisation (polychromatic model) increases the computational cost. For one iteration, we measure $6.7~\mathrm{min}$ on CPUs and $1.47~\mathrm{min}$ on one GPU; for two iterations, $11.3~\mathrm{min}$ (CPU) and $1.9~\mathrm{min}$ (GPU).

\paragraph{Non-local solver (multi-pixel shifts).}
In the strong-shift regime, the non-local solver is more robust but requires additional resampling work. With five spectral lines, one iteration requires $2~\mathrm{min}$ on one GPU for an oversampling factor of $2$, and about $8~\mathrm{min}$ for an oversampling factor of $4$.

\paragraph{Comparison with the original EPR implementation.}
For the same dataset size and geometry the original EPR benchmark reported on the order of one and a half days on 12 NVIDIA A40 GPUs for a dataset corresponding to approximately $\sim 30$ comparable volumes and polychromatic treatment; this yields an effective cost of $\sim 72$~min per volume on 12 GPUs, essentially all consumed by the phase retrieval process. Compared with the original EPR benchmark reported in \cite{Mirone2025JSR}, the present implementation reduces the phase-retrieval wall time per comparable volume by a factor of $580$ when normalised to the same GPU resources (about $2.8$ orders of magnitude) for one iteration with five-line spectral discretisation, enabling routine use in high-throughput workflows.

\section{Methods}

In this section we summarise the forward operators and the inversion scheme used in the implementation, and we provide the final expressions required to reproduce the algorithm.
The full saddle-point/WKB derivation of the local $O(L^2)$ intensity expansion is provided in the Supporting Information.

\subsection{Local forward model summary: $O(L^2)$ closure with WKB terminology}

We work with transverse coordinates $\boldsymbol r=(x,y)$ in the object plane ($z=0$) and $\boldsymbol r'=(x',y')$ in the detector plane ($z=L$). The scalar complex field is written as
$U(\boldsymbol r,0)=A(\boldsymbol r)\,e^{i\phi(\boldsymbol r)}$,
with intensity $I(\boldsymbol r,0)=|U(\boldsymbol r,0)|^2=A^2(\boldsymbol r)$.
We denote by $\nabla_{\!\perp}$ derivatives with respect to the transverse coordinates, and $k=2\pi/\lambda$ is the wavenumber.

We introduce $\varepsilon \equiv {L}/{k}$ and use a WKB expansion of the Fresnel propagator. In this hierarchy, WKB0 denotes the leading eikonal transport (ray map / intensity transport), while WKB1 denotes the first wave-optics correction (diffraction-pressure / quantum-potential term). The complete derivation is given in the Supporting Information; the resulting local intensity expansion reads
\begin{equation}
\begin{aligned}
I(\boldsymbol r',L)
\simeq\;&
I(\boldsymbol r',0)
-\varepsilon\,\partial_i\!\big(I\,\partial_i\phi\big)\Big|_{\boldsymbol r'}\\[4pt]
&+\frac{\varepsilon^2}{2}\,\partial_i\partial_j\!\big(I\,\partial_i\phi\,\partial_j\phi\big)\Big|_{\boldsymbol r'}
-\frac{\varepsilon^2}{4}\,\partial_i\!\big(I\,\partial_i Q\big)\Big|_{\boldsymbol r'}
+O(\varepsilon^3),
\end{aligned}
\label{eq-intensity3L}
\end{equation}
where
\[
Q(\boldsymbol r)\equiv \frac{\nabla_{\!\perp}^2\sqrt{I(\boldsymbol r,0)}}{\sqrt{I(\boldsymbol r,0)}}.
\]

While higher-order terms can in principle be derived, beyond $O(L^2)$ they increasingly encode wavefront/interference physics that cannot be represented fully within a local, single-valued WKB mapping. For this reason, we truncate at $O(L^2)$ (WKB0 plus the leading WKB1 contribution) in the present work.

\subsection{Non-local (multi-pixel) forward model: explicit WKB0 transport and discrete adjoint}

Throughout the Methods section, we denote by $\boldsymbol{\Delta}(\boldsymbol r)$ the physical eikonal-induced transverse shift, while $p$ denotes the detector pixel size.

\paragraph{Motivation and regime.}
The local second-order closure \Eq{eq-intensity3L} is obtained by Taylor-expanding WKB0 transport in powers of $\varepsilon=L/k$. This truncation is accurate when the associated eikonal ray shifts are typically sub-pixel. When shifts become comparable to, or larger than, the pixel size, the Taylor closure breaks down. To preserve robustness in this multi-pixel regime, we therefore replace the local Taylor closure by an explicit discrete realisation of the WKB0 mapping itself.

\paragraph{Per-line eikonal mapping and shift field.}
The WKB0 stationary-point condition defines the ray mapping
\begin{equation}
\boldsymbol r' = \boldsymbol r + \boldsymbol \Delta_s(\boldsymbol r),
\qquad
\boldsymbol \Delta_s(\boldsymbol r) \equiv \varepsilon_s\,\nabla_{\!\perp}\phi_s(\boldsymbol r),
\qquad
\varepsilon_s \equiv \frac{L}{k_s}.
\label{eq-nl-map}
\end{equation}
In practice we work in pixel units $\boldsymbol u_s(\boldsymbol r)=\boldsymbol \Delta_s(\boldsymbol r)/p$, where $p$ is the detector pixel size (or the internal pixel size on the oversampled grid).

\paragraph{Non-local forward operator as explicit WKB0 ray mapping (no Taylor expansion).}

The non-local forward model retains WKB0 forward mapping without expanding the delta:
\begin{equation}
P_s(\boldsymbol r')
=
\mathcal M_s[I_s](\boldsymbol r')
\equiv
\int_{\mathbb R^2} I_s(\boldsymbol r)\,
\delta\!\big(\boldsymbol r' - \boldsymbol r - \boldsymbol \Delta_s(\boldsymbol r)\big)\,d^2\boldsymbol r.
\label{eq-nl-forward}
\end{equation}
For a fixed shift field $\boldsymbol \Delta_s$, $\mathcal M_s$ is linear in $I_s$ and represents a mass-conserving transport. When $\|\boldsymbol u_s\|\ll 1$ pixel, the Taylor expansion of \Eq{eq-nl-forward} recovers the local eikonal series (and, when complemented by the second-order corrections, \Eq{eq-intensity3L}).

\paragraph{Discrete implementation: conservative bilinear redistribution on the detector grid.}
On a Cartesian detector grid, \Eq{eq-nl-forward} is evaluated by a conservative resampling: each source pixel at integer coordinate $(x,y)$ contributes its mass $I_s(x,y)$ to the mapped location
\[
(x_t,y_t) = (x,y) + \big(u_{x,s}(x,y),\,u_{y,s}(x,y)\big),
\]
which generally lies between pixels. The mass is distributed to the four surrounding detector pixels using bilinear weights. Denoting $x_0=\lfloor x_t\rfloor$, $y_0=\lfloor y_t\rfloor$, $d_x=x_t-x_0$, $d_y=y_t-y_0$, the weights are
\[
w_{00}=(1-d_x)(1-d_y),\quad
w_{10}=d_x(1-d_y),\quad
w_{01}=(1-d_x)d_y,\quad
w_{11}=d_x d_y,
\]
and the prediction is accumulated as
\begin{align*}
P_s(y_0,x_0)             &\mathrel{+}= I_s(y,x)\,w_{00},\\
P_s(y_0,x_0\!+\!1)       &\mathrel{+}= I_s(y,x)\,w_{10},\\
P_s(y_0\!+\!1,x_0)       &\mathrel{+}= I_s(y,x)\,w_{01},\\
P_s(y_0\!+\!1,x_0\!+\!1) &\mathrel{+}= I_s(y,x)\,w_{11}.
\end{align*}

\paragraph{Adjoint (transpose) operator: bilinear interpolation of the detector residual.}
With this definition, the back-transport of a detector residual is performed with the exact adjoint (transpose) of the discrete bilinear redistribution operator. In practice, this adjoint action is numerically equivalent to bilinear interpolation of the detector residual at the mapped location $(x_t,y_t)$.
\begin{equation}
\begin{aligned}
\big(\mathcal{M}_{s}^{\!*} e_s\big)(x,y)  &=  w_{00}\,e_s(y_0,x_0)
 + w_{10}\,e_s(y_0,x_0\!+\!1) \\
&\quad
 + w_{01}\,e_s(y_0\!+\!1,x_0)
 + w_{11}\,e_s(y_0\!+\!1,x_0\!+\!1),
\end{aligned}
\label{eqnladjoint}
\end{equation}
which is numerically equivalent to bilinear interpolation of $e_s$ at $(x_t,y_t)$,  but written in this explicit adjoint form to make the transpose relationship clear.

\subsection{Polychromatic framework: spectral discretisation (local and non-local)}

\paragraph{Polychromatic measurement as an incoherent spectral sum.}
We discretise the spectrum into $N_s$ effective spectral points (``lines'') indexed by $s=1,\dots,N_s$. The detector intensity is modelled as an incoherent sum of the propagated intensities of the spectral components:
\begin{equation}
I_{\mathrm{det}}(\boldsymbol r)
\approx
\sum_{s=1}^{N_s}\,\mathcal F_{s}\!\big[I_s\big](\boldsymbol r),
\qquad
\mathcal F_s\in\{\mathcal P_{m,s},\,\mathcal M_s\}.
\label{eq-poly-forward}
\end{equation}
In the local formulation we take $\mathcal F_s=\mathcal P_{m,s}$, i.e.\ the Taylor/WKB closure truncated at order $m\in\{1,2\}$. In the non-local formulation we take $\mathcal F_s=\mathcal M_s$, i.e.\ the explicit WKB0 transport operator \Eq{eq-nl-forward}.

\paragraph{Single-material phase--absorption relation (per spectral line).}
For a single-material (homogeneous) object with refractive index $n=1-\delta+i\beta$, both phase and attenuation are proportional to the projected thickness, and eliminating thickness yields
\begin{equation}
\phi_s(\boldsymbol r)
=
\frac{\delta_s}{2\beta_s}\,\ln I_s(\boldsymbol r)
=
\frac{\mathrm{db}_s(x)}{2}\,\ln I_s(\boldsymbol r)
\qquad
(\text{up to an additive constant}),
\label{eq-poly-phase-lnI}
\end{equation}
where $\mathrm{db}_s(x)=\delta_s(x)/\beta_s(x)$ may be $x$-dependent if an $x$-dependent effective spectrum is used.
Operationally, this relation is used to express $\nabla_{\!\perp}\phi_s$ in terms of $\nabla_{\!\perp}\ln I_s$ only.

\paragraph{Remark on spatially varying spectra.}
The implementation supports slow spatial variations of the effective spectrum across the detector (e.g.\ due to filtration, container thickness, or source/optics non-uniformity). If such variations are negligible, one can use a spatially uniform spectrum. Detailed file format and the blockwise interpolation strategy are provided in the Supporting Information.

\subsection{Iterative inversion (local and non-local)}

\paragraph{Unknowns and single-thickness constraint.}
In the polychromatic model \Eq{eq-poly-forward}, we represent the object-plane transmission as a set of nonnegative spectral components $\{I_s(\boldsymbol r)\}_{s=1}^{N_s}$ (each $I_s$ already includes its local spectral weight). Under the homogeneous-composition/single-material hypothesis, these components are constrained by a single nominal thickness field. We enforce this by parameterising
\begin{equation}
I_s(\boldsymbol r)
=
f_s(x)\,\exp\!\big(-\gamma_s(x)\,\tau(\boldsymbol r)\big),
\qquad s=1,\dots,N_s,
\label{eq-poly-singlethickness}
\end{equation}
with $f_s(x)$ the local spectral fraction and $\gamma_s(x)$ the per-line attenuation scaling.

\paragraph{Iteration structure.}
At iteration $n$, given $\{I_s^{(n)}\}$, we compute per-line predictions and their sum, form the detector residual, split it across spectral lines with positivity-weighted weights, transport the per-line residuals back to the object grid (identity in the local solver, and the corresponding adjoint (transpose) operator \Eq{eqnladjoint} in the non-local solver), and apply a per-line FFT-diagonal inverse of the linear operator (Paganin-type) as a preconditioner update.

\paragraph{Projection to the single-thickness manifold.}
After each iteration (for $N_s>1$), we project the updated components back to the single-thickness form \Eq{eq-poly-singlethickness}. This stabilises the inversion and prevents drift into non-physical degrees of freedom. Implementation details are given in the Supporting Information.

\section{Conclusion}

We have presented an accelerated second-order formulation of eikonal phase retrieval for near-field propagation-based imaging, obtained by retaining the complete $O(L^{2})$ content of a saddle-point (WKB) expansion of the Fresnel propagator. The resulting forward model combines the quadratic eikonal transport contribution (WKB0) with the leading wave-optics correction (diffraction-pressure, WKB1), while remaining expressible through transverse derivatives evaluated in the object plane. This yields an efficient iterative scheme naturally preconditioned by FFT-diagonal single-distance inverse operators.

To remain robust beyond the sub-pixel regime, we complemented the local $O(L^{2})$ closure with a non-local mapping-based solver that explicitly
 transports intensity by a mass-conserving forward remapping and transfers detector residuals back to the object grid with the corresponding discrete adjoint (transpose) operator 
  before applying the  $\mathcal{P}_{1,s}^{-1}$ preconditioner. The two solvers therefore form a consistent bridge between regimes where refraction-induced shifts are small and regimes where multi-pixel transport dominates.

The same framework supports a practical polychromatic treatment by discretising the spectrum into a finite set of effective spectral lines and summing their propagated contributions. In the HiP-CT context, this improves agreement with experimental data by accounting for spectral effects and thereby reducing residual bias and artefacts that cannot be captured by a single effective energy.

Compared with the original EPR implementation reported in \cite{Mirone2025JSR}, the present approach reduces the phase-retrieval wall time per comparable volume by approximately a factor of $580$ when normalised to the same GPU resources (about $2.8$ orders of magnitude), making nonlinear phase retrieval compatible with routine, high-throughput reconstruction workflows.

\paragraph{Practical impact and integration in reconstruction workflows.}
In the current \texttt{NightRail} workflow, phase retrieval and tomographic reconstruction are scheduled to exploit hardware concurrency. When a single GPU is available, phase retrieval can be executed on CPUs while the GPU performs back-projections. For large acquisitions processed in chunks, the workflow overlaps computation: while the GPU back-projects the current chunk, the CPUs pre-process the next chunk, including the phase-retrieval step. When two GPUs are available and the EPR-WKB option is active, one GPU can be reserved for phase retrieval while the other GPU concurrently executes the back-projections. Because the algorithmic complexity of backprojection is higher than that of FFT-based filtering, the additional time required by EPR-WKB can often remain hidden behind tomographic reconstruction in practical high-throughput settings.

Beyond the specific HiP-CT use case, fast access to a second-order near-field model with an explicit wave-optics correction opens the possibility to explore propagation distances closer to the near-field boundary (and potentially slightly beyond it), where standard first-order, linearised single-distance approaches become unreliable. This operating space is particularly relevant for applications pushing towards stronger phase gradients, such as lower-energy imaging and/or higher spatial resolution at modern synchrotron sources, where improved artefact control and quantitative robustness are critical.

\section{Funding.}
\begin{itemize}
\item ESRF funding proposals md1290 and md1389, performed at beamlines BM18 and BM05.
\item This work was supported in part by the Chan Zuckerberg Initiative DAF (grant 2022-316777).
\end{itemize}

\section{Disclosures.}
The authors declare no conflicts of interest.

\section{Data Availability}
The code sources to reproduce the shown images and a data subset can be retrieved as detailed in the Supporting Information.

\bibliographystyle{iucr}
\bibliography{eprwkb}

\clearpage

\pagebreak

\section{Supporting Information}

\setcounter{equation}{0}
\renewcommand{\theequation}{S\arabic{equation}}
\setcounter{figure}{0}
\renewcommand{\thefigure}{S\arabic{figure}}
\setcounter{table}{0}
\renewcommand{\thetable}{S\arabic{table}}

\subsection{Supplementary Methods: Fresnel propagation, WKB expansion, and the diffraction-pressure term}
\label{sec:supp_methods}

This Supporting Information provides the detailed derivations that underpin the forward models and terminology
used in the main text (WKB0/WKB1, local $O(L^2)$ closure, and the diffraction-pressure term).
The goal is twofold: (i) to make the mathematical chain of approximations explicit (from Helmholtz to Fresnel to WKB),
and (ii) to clarify the physical meaning of the WKB1 contribution as the leading wave-optics correction
within an otherwise eikonal, point-to-point mapping framework.

\subsubsection{From the Helmholtz equation to the Fresnel propagator as a Green's function}
\label{sec:helmholtz_to_fresnel}

\paragraph{Scalar Helmholtz equation and paraxial reduction.}
We start from the monochromatic scalar Helmholtz equation for a field $U(x,y,z)$ in free space,
\begin{equation}
\left(\nabla_{\!\perp}^2 + \partial_{zz} + k^2\right) U(x,y,z)=0,
\label{eqS-helmholtz}
\end{equation}
where $\nabla_{\!\perp}^2=\partial_{xx}+\partial_{yy}$ is the transverse Laplacian and $k=2\pi/\lambda$.
Introducing the standard slowly varying envelope $\psi$ through
\begin{equation}
U(x,y,z)=e^{ikz}\,\psi(x,y,z),
\label{eqS-envelope}
\end{equation}
and substituting into \Eq{eqS-helmholtz} yields
\[
\nabla_{\!\perp}^2 \psi + \partial_{zz}\psi + 2ik\,\partial_z \psi = 0.
\]
Under the paraxial approximation (slow longitudinal variation of the envelope), $\partial_{zz}\psi$ is neglected
compared with $2ik\,\partial_z\psi$, leading to the paraxial wave equation
\begin{equation}
2ik\,\partial_z \psi + \nabla_{\!\perp}^2 \psi = 0.
\label{eqS-paraxial}
\end{equation}
Equation \Eq{eqS-paraxial} is equivalent to a Schr\"odinger-type evolution in the propagation variable $z$,
\begin{equation}
i\,\partial_z \psi = -\frac{1}{2k}\,\nabla_{\!\perp}^2 \psi.
\label{eqS-schroedinger}
\end{equation}

\paragraph{Green's function and Fresnel integral.}
The solution of \Eq{eqS-paraxial} with initial condition $\psi(\boldsymbol r,0)$ can be written using the
Green's function (propagator) $G(\boldsymbol r'-\boldsymbol r;L)$ satisfying the same equation and a
$\delta$-source at $z=0$:
\[
\psi(\boldsymbol r',L)=\int_{\mathbb R^2} G(\boldsymbol r'-\boldsymbol r;L)\,\psi(\boldsymbol r,0)\,d^2\boldsymbol r.
\]
The free-space paraxial Green's function is
\begin{equation}
G(\boldsymbol \rho;L) = \frac{k}{2\pi i L}\,
\exp\!\left(i\frac{k}{2L}\,|\boldsymbol \rho|^2\right),
\qquad \boldsymbol \rho=\boldsymbol r'-\boldsymbol r,
\label{eqS-green}
\end{equation}
so that
\begin{equation}
\psi(\boldsymbol r',L)
=
\frac{k}{2\pi i L}
\int_{\mathbb R^2}
\psi(\boldsymbol r,0)\,
\exp\!\left(i\frac{k}{2L}\,|\boldsymbol r' - \boldsymbol r|^2\right)\,
d^2\boldsymbol r.
\label{eqS-fresnel-psi}
\end{equation}
Recovering $U=e^{ikz}\psi$ gives the standard Fresnel (paraxial Huygens--Fresnel) integral
\begin{equation}
U(\boldsymbol r',L)
=
\frac{k}{2\pi i L}\,e^{ikL}
\int_{\mathbb R^2}
U(\boldsymbol r,0)\,
\exp\!\left(i\frac{k}{2L}\,|\boldsymbol r' - \boldsymbol r|^2\right)\,
d^2\boldsymbol r,
\label{eq-hf}
\end{equation}
which is therefore the Green's function solution of the paraxial reduction of the Helmholtz equation.
This observation is useful because it allows the WKB corrections to be interpreted both from the
stationary-phase expansion of \Eq{eq-hf} and from the hydrodynamic (Madelung) form of the paraxial PDE.

\subsubsection{Saddle-point (WKB) expansion of the Fresnel integral}
\label{sec:wkb_saddle}

We work with transverse coordinates $\boldsymbol r=(x,y)$ in the object plane ($z=0$) and
$\boldsymbol r'=(x',y')$ in the detector plane ($z=L$). We write the object-plane field as
$U(\boldsymbol r,0)=A(\boldsymbol r)\,e^{i\phi(\boldsymbol r)}$, with intensity
$I(\boldsymbol r,0)=|U(\boldsymbol r,0)|^2=A^2(\boldsymbol r)$.
For bookkeeping we introduce
\begin{equation}
\varepsilon \equiv \frac{L}{k} = \frac{\lambda L}{2\pi},
\label{eqS-eps}
\end{equation}
which has units of length$^2$.
The near-field condition corresponds to a large Fresnel number $N_F=\ell^2/(\lambda L)\gg 1$, where $\ell$
is a characteristic transverse length scale of variation of $A$ and $\phi$.

\paragraph{WKB0 saddle: eikonal ray map.}
To put \Eq{eq-hf} into stationary-phase form, we collect the free-space quadratic phase and the object phase
into the phase function
\begin{equation}
\Phi(\boldsymbol r;\boldsymbol r')
\equiv
\frac{1}{2}\,|\boldsymbol r' - \boldsymbol r|^2
+
\varepsilon\,\phi(\boldsymbol r),
\qquad
i\phi + i\frac{k}{2L}|\boldsymbol r'-\boldsymbol r|^2
=
\frac{i}{\varepsilon}\,\Phi.
\label{eqS-Phi}
\end{equation}
Stationary points $\boldsymbol r_s$ satisfy $\partial_i\Phi(\boldsymbol r_s;\boldsymbol r')=0$, giving the
near-identity eikonal map
\begin{equation}
\boldsymbol r'=\boldsymbol r_s+\varepsilon\,\nabla_{\!\perp}\phi(\boldsymbol r_s).
\label{eq-raymap0}
\end{equation}
In the near-field regime $\varepsilon\nabla_{\!\perp}\phi$ is small so the mapping remains single-valued
(no caustics) and the saddle lies close to $\boldsymbol r_s\simeq \boldsymbol r'$.

Expanding $\Phi$ about the saddle by writing $\boldsymbol r=\boldsymbol r_s+\boldsymbol u$ yields
\begin{equation}
\Phi(\boldsymbol r_s+\boldsymbol u;\boldsymbol r')
\simeq
\Phi_s+\frac{1}{2}\,u_i H_{ij}u_j,
\qquad
H_{ij}=\delta_{ij}+\varepsilon(\partial_i\partial_j\phi)_s ,
\label{eq-hess}
\end{equation}
with $\Phi_s=\Phi(\boldsymbol r_s;\boldsymbol r')$.
Keeping the amplitude fixed at $A_s=A(\boldsymbol r_s)$ and evaluating the Gaussian integral yields the
leading WKB0 field
\begin{equation}
U_{\mathrm{WKB0}}(\boldsymbol r',L)
=
e^{ikL}\,\frac{A_s}{\sqrt{\det H}}\,
\exp\!\left(\frac{i}{\varepsilon}\,\Phi_s\right).
\label{eq-uwkb0}
\end{equation}

\paragraph{WKB1 saddle: amplitude curvature and the ``diffraction-pressure'' phase.}
The first correction arises by retaining the local curvature of $A=\sqrt{I}$ around the saddle,
\begin{equation}
A(\boldsymbol r_s+\boldsymbol u)
=
A_s
+
(\partial_i A)_s\,u_i
+
\frac{1}{2}(\partial_i\partial_j A)_s\,u_i u_j
+
O(|\boldsymbol u|^3).
\label{eq-amp}
\end{equation}
The linear term integrates to zero by symmetry, and the quadratic term yields
\begin{equation}
U(\boldsymbol r',L)
\simeq
U_{\mathrm{WKB0}}(\boldsymbol r',L)
\left[
1+\frac{i\varepsilon}{2}\,
\frac{(\partial_i\partial_j A)_s (H^{-1})_{ij}}{A_s}
\right].
\label{eq-uwkb1lin}
\end{equation}
In the near-field regime $H^{-1}=\mathbf I_2+O(\varepsilon)$, so the correction reduces to the local curvature
ratio
\begin{equation}
Q(\boldsymbol r)\equiv \frac{\nabla_{\!\perp}^2 A(\boldsymbol r)}{A(\boldsymbol r)}
=
\frac{\nabla_{\!\perp}^2\sqrt{I(\boldsymbol r,0)}}{\sqrt{I(\boldsymbol r,0)}}.
\label{eq-qdef}
\end{equation}
This term acts as a local phase correction to the field. Its leading impact on the propagated intensity
appears at order $O(\varepsilon^2)$.

\paragraph{WKB0 forward model and local intensity expansion.}
Neglecting multi-branch interference (single relevant saddle), WKB0 transport can be written as a mass-conserving mapping of
intensity along the ray map \Eq{eq-raymap0}:
\begin{equation}
I(\boldsymbol r',L)=\int I(\boldsymbol r,0)\,
\delta\!\left(\boldsymbol r'-\boldsymbol r-\varepsilon\nabla_{\!\perp}\phi(\boldsymbol r)\right)\,d^2\boldsymbol r .
\label{eq-push0}
\end{equation}
Expanding the delta distribution in $\varepsilon$ and integrating by parts gives the WKB0 (eikonal) series
\begin{equation}
I(\boldsymbol r',L)
\simeq
I(\boldsymbol r',0)
-\varepsilon\,\partial_i\!\big(I\,\partial_i\phi\big)\Big|_{\boldsymbol r'}
+\frac{\varepsilon^2}{2}\,\partial_i\partial_j\!\big(I\,\partial_i\phi\,\partial_j\phi\big)\Big|_{\boldsymbol r'}
+O(\varepsilon^3).
\label{eq-int-eik}
\end{equation}

\subsubsection{Diffraction-pressure term from the Madelung form of the paraxial equation}
\label{sec:madelung_qpressure}

\paragraph{Hydrodynamic (Madelung) representation.}
Because Fresnel propagation is the Green's function solution of the paraxial equation \Eq{eqS-paraxial},
one can equivalently derive WKB corrections by rewriting \Eq{eqS-schroedinger} in amplitude-phase variables.
Let
\[
\psi(\boldsymbol r,z)=\sqrt{I(\boldsymbol r,z)}\,e^{i\phi(\boldsymbol r,z)}.
\]
Substituting into \Eq{eqS-schroedinger} and separating real and imaginary parts yields the exact pair
\begin{equation}
\partial_z I+\frac{1}{k}\,\nabla_{\!\perp}\!\cdot\!\big(I\nabla_{\!\perp}\phi\big)=0,
\label{eqS-continuity}
\end{equation}
\begin{equation}
\partial_z\phi+\frac{1}{2k}\,|\nabla_{\!\perp}\phi|^2-\frac{1}{2k}\,Q=0,
\label{eqS-hj}
\end{equation}
with $Q$ defined in \Eq{eq-qdef}.
The $Q$-term is the well-known quantum-potential (quantum-pressure) contribution in the Madelung
formulation~\cite{Madelung1927}, and in the present context it is naturally interpreted as a local
``diffraction-pressure'' correction to purely geometrical transport.

\paragraph{Intensity expansion to $O(\varepsilon^2)$ (WKB0+WKB1).}
Taylor-expanding
\[
I(\boldsymbol r',L)=I(\boldsymbol r',0)+L\,\partial_z I|_{0}+\frac{L^2}{2}\,\partial_z^2 I|_{0}+O(L^3),
\]
and using \Eq{eqS-continuity}--\Eq{eqS-hj} to eliminate $\partial_z I$ and $\partial_z\phi$ yields an additional
$O(\varepsilon^2)$ contribution to the intensity series, namely
$-(\varepsilon^2/4)\,\partial_i(I\,\partial_i Q)$.
Combining this with the WKB0 transport terms gives the local WKB0+WKB1 intensity expansion used in the main text:
\begin{equation}
\begin{aligned}
I(\boldsymbol r',L)
\simeq\;&
I(\boldsymbol r',0)
-\varepsilon\,\partial_i\!\big(I\,\partial_i\phi\big)\Big|_{\boldsymbol r'}\\[4pt]
&+\frac{\varepsilon^2}{2}\,\partial_i\partial_j\!\big(I\,\partial_i\phi\,\partial_j\phi\big)\Big|_{\boldsymbol r'}
-\frac{\varepsilon^2}{4}\,\partial_i\!\big(I\,\partial_i Q\big)\Big|_{\boldsymbol r'}
+O(\varepsilon^3).
\end{aligned}
\label{eqS-intensity3L}
\end{equation}
This derivation emphasizes that the diffraction-pressure term is not an additional model assumption:
it is an exact component of the paraxial wave equation expressed in amplitude-phase variables, and it
enters the local near-field intensity expansion at the same order as the quadratic eikonal transport
correction.

\paragraph{Scope and interpretation of WKB1.}
In our formulation, WKB0 corresponds to a point-to-point transport map (ray mapping), while WKB1 provides
the leading local wave-optics correction compatible with this mapping picture.
As one moves further away from the near-field regime, increasingly important physics involves wavefront
structure and interference that cannot be fully represented within a single-valued, local mapping framework.
This motivates retaining WKB0 and including only the leading WKB1 term at $O(L^2)$ as a compact estimate of
wave-optics effects while staying within the eikonal-WKB setting.

\subsection{Supplementary Note: spatially varying effective spectrum (example and general applicability)}
\label{sec:supp_spectrum}

This note provides additional context on the $x$-dependent effective spectrum parameterization used in the
polychromatic implementation. The same formalism applies to other beamlines and configurations whenever the
transmitted spectrum varies across the field of view (e.g. due to filtration, sample container thickness,
or source/optics non-uniformity). If such variations are negligible, the model reduces to a spatially uniform
spectrum (single block).

\paragraph{Why an $x$-dependent spectrum can arise.}
In some PPC-$\mu$CT configurations, specimens are immersed in a cylindrical container whose axis is aligned
with the vertical detector direction. In such cases, upstream filtration introduced by the container (and by
the specimen itself) varies predominantly across the horizontal coordinate $x$: rays crossing the jar center
traverse a longer path length in the container material than rays closer to the lateral sides. For sufficiently
large containers, this alone can produce a marked spatial variation of the transmitted effective spectrum
(beam hardening stronger at the center and weaker towards the edges).

\paragraph{Example: BM18 HiP-CT.}
On BM18, the incident spectrum can also be non-uniform across $x$ because of the tripole wiggler design:
the final beam results from the superimposition of emission from a central pole (higher field) and two lateral
poles (lower field). In the HiP-CT operating regime (around the equilibrium energy, about $100$ keV), these effects
combine in a non-trivial way, motivating an $x$-dependent effective spectrum parameterization.

\paragraph{Extensions.}
A dependence on the vertical coordinate $y$ is straightforward to include if required (e.g. by tabulating
two-dimensional spectral maps instead of one-dimensional profiles). In many helical acquisition settings,
a purely $x$-dependent model is already a robust compromise because a detector row is not associated with a unique
physical height throughout the scan.

\subsection{Data availability and reproducibility (NightRail)}
\label{sec:supp_repro}

\subsubsection{Demo dataset and NightRail input files}
All examples shown in this paper (including the demo dataset and the corresponding NightRail input
configuration files) are available from the ESRF data repository at:
\begin{center}
\texttt{https://cloud.esrf.fr/s/AF59f7E393kDiTk}
\end{center}
The archive contains a ready-to-run \texttt{RECONSTRUCTIONS} directory with multiple reconstruction
cases matching the configurations discussed in the manuscript.

\subsubsection{Installing the NightRail BM18 workflow and running the examples}
In addition to the demo archive above, reproducing the results requires installing the NightRail BM18
application code. The installation is performed in two steps.

\paragraph{Step 1: clone the repository.}
Clone the repository and check out the \texttt{eprwkb} branch:
\begin{verbatim}
git clone https://gitlab.esrf.fr/night_rail/applications/mirone/night_rail_bm18
cd night_rail_bm18
git checkout eprwkb
\end{verbatim}

\paragraph{Step 2: create the Python virtual environment and install the code.}
From the cloned directory, enter \texttt{install\_u24\_venv} where you find the installation script.

\paragraph{Step 3: adapt the installation script.}
Edit the script
\begin{verbatim}
no_surprise_installation_script_u24.sh
\end{verbatim}
to select the desired target directory for the virtual environment. In particular, set:
\begin{verbatim}
TAG=eprwkb
\end{verbatim}
and run the script to complete the installation.

\subsubsection{Running an example reconstruction}
All demo cases are prepared ready to run without modification. As a representative example, consider the directory
\texttt{RECONSTRUCTIONS/epr\_mono\_3iter}. From the demo directory:
\begin{verbatim}
cd RECONSTRUCTIONS/epr_mono_3iter
\end{verbatim}

\paragraph{Generate the workflow scripts (graph expansion).}
\begin{verbatim}
night_rail_bm18_create_scripts -i nrbm18_default.json
\end{verbatim}

\paragraph{Run the workflow scripts.}
Then execute the generated session script:
\begin{verbatim}
./unnamed_session_run_graph.sh
\end{verbatim}

\end{document}